\documentclass[prl,twocolumn,showpacs,preprintnumbers,amsmath,amssymb]{revtex4}
\usepackage[]{graphicx}
\usepackage[]{textcomp}

\newcommand{\ket}[1]{\mbox{\ensuremath{|#1\rangle}}}
\newcommand{\bra}[1]{\mbox{\ensuremath{\langle#1|}}}
\begin{document}
\begin{abstract}
\end{abstract}
\title{Beating Abbe diffraction limit in confocal microscopy via non-classical photon statistics
}
\author{D. Gatto Monticone$^{1,2,3}$, K. Katamadze$^{4,5}$, P. Traina$^6$, E. Moreva$^{6,7}$, J. Forneris$^{1,2,3}$, I. Ruo-Berchera$^6$, P. Olivero$^{1,2,3}$, I. P. Degiovanni$^6$, G. Brida$^6$ and M. Genovese$^{3,6}$.}
\affiliation{$^1$Physics Department and NIS Inter-departmental Centre - University of Torino, Torino, Italy}
\affiliation{$^2$Istituto Nazionale di Fisica Nucleare (INFN) Sez. Torino, Torino, Italy}
\affiliation{$^3$Consorzio Nazionale Interuniversitario per le Scienze Fisiche della Materia (CNISM) Sez. Torino, Torino, Italy}
\affiliation{$^4$M. V. Lomonosov Moscow State University, Moscow, Russia}
\affiliation{$^5$Russian Academy of Sciences, Institute of Physics and Technology, Moscow, Russia}
\affiliation{$^6$Istituto Nazionale di Ricerca Metrologica (INRiM), Torino, Italy}
\affiliation{$^7$International Laser Center of M.V.Lomonosov Moscow State University, Moscow, Russia}


\begin{abstract}
 We experimentally demonstrate quantum enhanced resolution in confocal fluorescence microscopy exploiting the non-classical photon statistics of single nitrogen-vacancy colour centres in diamond. By developing a general model of super-resolution based on the direct sampling of the $k^{th}$-order autocorrelation function of the photoluminescence signal, we show the possibility to resolve in principle arbitrarily close emitting centres.
 \end{abstract}

\maketitle



In the last decade, measurement techniques enhanced by using peculiar properties of quantum light  \cite{maccone,Banaszek}
has been successfully demonstrated in several remarkable real application scenarios, for example interferometric measurements aimed to reveal gravitational waves and quantum gravity effect\cite{Abadie,holo}, biological particles tracking \cite{Taylor}, phase contrast microscopy \cite{Ono}, and imaging \cite{ivano,brida}.

Very recently, a novel technique to beat the diffraction limit in microscopy that relies on the anti-bunching behaviour of photons emitted by single fluophores has been proposed \cite{sresth}, and realized in wide field microscopy \cite{sresth2} by using an EMCCD camera.The maximum obtainable imaging resolution in classical far-field fluorescence microscopy, according to the Abbe diffraction limit, is $R\simeq\lambda/{2NA}$, where $\lambda$ is the wavelength of the light and $NA$ is the numerical aperture of the objective.
This restricts the current capability of precisely measuring the position of very small objects such as  single photon emitters (colour centres, quantum dots, etc.) \cite{emitter1,emitter2,emitter3,emitter4,emitter5,SiV,njp,alan,alan2}, limiting their potential exploitation in the frame quantum technology \cite{jelezko1,jelezko2}. In general,  the research of methods to obtain a microscopy resolution below the diffraction limit is a topic of the utmost interest \cite{kok,thiel,kok2,geno1,geno2,geno3,geno4,geno5} that could provide dramatic improvement in the observation of several systems spanning from quantum dots \cite{masha} to living cells \cite{cells1,cells2,cells3,cells4}. As a notable example, in several entanglement-related experiments using strongly coupled-single photon emitters it is of the utmost importance to measure their positions with the highest spatial resolution \cite{needref}.
In principle this limitation can be overcome by recently developed microscopy techniques such as  Stimulated Emission Depletion (STED) and Ground State Depletion (GSD) \cite{sted,gsd}. Nevertheless, even if they have been demonstrated effectively able to provide super-resolved imaging in many specific applications, among which colour centres in diamond \cite{colcen}, they are characterized by rather specific experimental requirements (dual laser excitation system, availability of luminescence quenching mechanisms by stimulated emission, non-trivial shaping of the quenching beam, high power). Furthermore, these techniques are not suitable in applications in which the fluorescence is not optically induced \cite{electro, electro2}, so that new methods are required for those applications.

Inspired by the works in Ref.s \cite{sresth}, in this letter we study in general the possibility to beating the diffraction limit by using high order Glauber correlation functions $g^{(k)}(t=0)$, showing that the knowledge of the spatial map of the correlation up to $k-$th order, together with the intensity map, allows a $1/\sqrt{k}$ corresponding improvement of resolution. It turns out that in some cases, when it is reasonable to assume $g^{(k)}=0$ for $k>k_{0}$, just measuring $g^{(2)},..,g^{(k_{0})}$, allows in principle to approach an arbitrary resolution.
We experimentally test super-resolution in the significant case of confocal microscopy for the first time, considering clusters of few NV centres in artificial diamond grown by Chemical Vapour Deposition (CVD) and using a detector-tree of commercial (non-photon-number-resolving) single-photon detectors \cite{tree,alan} .
We demonstrate a resolution increasing by sampling the $g^{(2)}$ of the signal, and a further improvement by measuring  $g^{(3)}$. Furthermore, we show that just by considering the contribution of higher powers of $g^{(2)}$, when only two centres are relevant (as certified by $g^{(3)}=0$), larger improvement in the resolution can be obtained, as predicted by the theory.
This technique appears particularly valuable since the sampling of $g^{(2)}$ is a widely used 
and well established experimental procedure to test 
the statistical 
properties of 
quantum optical sources in general, and of single-photon sources in paricular, thus its adoption may come at almost zero cost. 



Let $\mathcal{P}(x)$  be the probability of detecting a photon at the image position $x$ from a single photon emitter upon a pulsed excitation \cite{footnote1}.

The function $\mathcal{P}(x)$ is typically an unimodal distribution and when fluorescence saturation effects are neglected the normalized $\mathcal{P}(x)$ represents the Point Spread Function (PSF) of the microscope.  In general when taking the $k^{th}$ power, $[\mathcal{P}(x)]^k$, the function gets narrower. In most cases $\mathcal{P}(x)$ can be well fitted by a Gaussian function, so the full-width-at-half-maximum (FWHM) of  $[\mathcal{P}(x)]^k$ reduces by a factor $\sqrt{k}$.

The fluorescence signal $S$ coming from $n$ arbitrarily distributed emitters  in a specific image position $x$ is then proportional to
\begin{equation}
S(x)\propto\sum_{\alpha=1}^n \mathcal{P_\alpha}(x).
\end{equation}
In order to obtain a resolution enhancement in the case of the $n$ single-photon emitters, i.e. to resolve the presence and the position of two or more centers when they are closer than the PSF, it would be useful to have a function containing pure $k^{th}$ powers of each single emitter probability $\sum_{\alpha=1}^n [\mathcal{P_\alpha}(x)]^k$. Unfortunately,
the $k^{th}$ power of the signal itself contains also the cross products terms ($c. p.$) multiplied by the appropriate multinomial coefficients,
\begin{equation}
S^k(x)\propto\sum_{\alpha=1}^n [\mathcal{P_\alpha}(x)]^k+c.p.
\end{equation}

The method described in the following allows the removal of the contribution from cross product terms using photon correlations, resulting in an effective increase of optical resolution.

To simplify our analysis and without any loss of generality,  we consider that all the losses and inefficiencies are accounted for as part of the source the  quantum state, i.e. the state $\hat{\rho}_\alpha(x)$ corresponding to the single photon emission of the $\alpha^{th}$ source that is detected at the position $x$ of the of an ideal (unit-quantum efficiency and photon-number-resolving) single-photon detector can be expressed as:
\begin{equation}
\hat{\rho}_\alpha(x)=\mathcal{P_\alpha}(x)\ket{1}_\alpha\bra{1}_\alpha+[1-\mathcal{P_\alpha}(x)]\ket{0}_\alpha\bra{0}_\alpha.
\end{equation}

The multi-photon state generated by several $n$ single-photon emitters is then:
\begin{equation}
\hat{\rho}=\bigotimes_{\alpha=1}^n\hat{\rho}_\alpha.
\end{equation}

Defining the number of detected photons from the system of single-photon emitters $\hat{N}=\sum_{\alpha=1}^n \hat{a}_\alpha^\dagger \hat{a}_\alpha$, one obtains that:
\begin{equation}
\langle \hat{N} \rangle=tr[\hat{\rho}\hat{N}]=\sum_{\alpha=1}^n\mathcal{P_\alpha}(x).
\end{equation}

In this case we define the $k^{th}$-order auto-correlation function as:
\begin{equation}
g^{(k)}=\frac{\langle\prod_{i=0}^{k-1}(\hat{N}-i)\rangle}{\langle \hat{N} \rangle^k} .
\end{equation}
(Note that in case of CW excitation the functions have time dependence: we are dealing with the value of $g^{(k)}(t = 0)$)
Knowing the value of $\langle\hat{N}\rangle$ and the set of $g^{(i)}$ ($1\leq i\leq k$), an image with increased resolution can be ideally obtained at any order $k$. For instance, the expressions of the super-resolved images for orders spanning from $k = 2$ to $k = 5$ are:

\begin{equation}
\label{eq:k2}
\sum_{\alpha=1}^2 [\mathcal{P_\alpha}(x))]^2=\langle\hat{N}\rangle^2[1-g^{(2)})],
\end{equation}

\begin{equation}
\label{eq:k3}
\sum_{\alpha=1}^3 [\mathcal{P_\alpha}(x)]^3=\langle\hat{N}\rangle^3[1-\frac{3}{2}g^{(2)}+\frac{1}{2}g^{(3)})],
\end{equation}

\begin{equation}
\label{eq:k4}
\sum_{\alpha=1}^4 [\mathcal{P_\alpha}(x)]^4=\langle\hat{N}\rangle^4\{1-2g^{(2)}+\frac{1}{2}[g^{(2)}]^2+\frac{2}{3}g^{(3)}-\frac{1}{6}g^{(4)}\},
\end{equation}

\begin{align}
\label{eq:k5}
\sum_{\alpha=1}^5 [\mathcal{P_\alpha}(x))]^5=\langle\hat{N}\rangle^5\{1-\frac{5}{2}g^{(2)}+\frac{5}{4}[g^{(2)}]^2+\frac{5}{6}g^{(3)}+\nonumber \\
-\frac{5}{12}g^{(2)}g^{(3)}-\frac{5}{24}g^{(4)}+\frac{1}{24}g^{(5)}\}.
\end{align}

In general, the expressions of the super-resolved images for any $k$ have the following form:
\begin{equation}
\label{eq:k8}
\sum_{\alpha=1}^n [\mathcal{P_\alpha}(x)]^k=\langle\hat{N}\rangle^k\sum_{i=1}^{i_{max}}y_i \beta_{i},
\end{equation}
where  $\beta_{i}$ represent in general products of the form $g^{(j_1)} \cdot g^{(j_2)} \cdot \ldots \cdot g^{(j_l)}$, $i_{max}$ being the number of possible (ordered) combinations, satisfying the condition $\sum_{p=1}^l j_p=k$ ($g^{(1)}$ is equal to $1$ according to Eq. (7)) 
 and $y_i$ are multiplicative coefficients that can be straightforwardly calculated, as it is shown in the cases up to $k=5$ in Eqs. (8)-(11).



\begin{figure}
   \begin{center}
   \begin{tabular}{c}
   \includegraphics[scale=0.4]{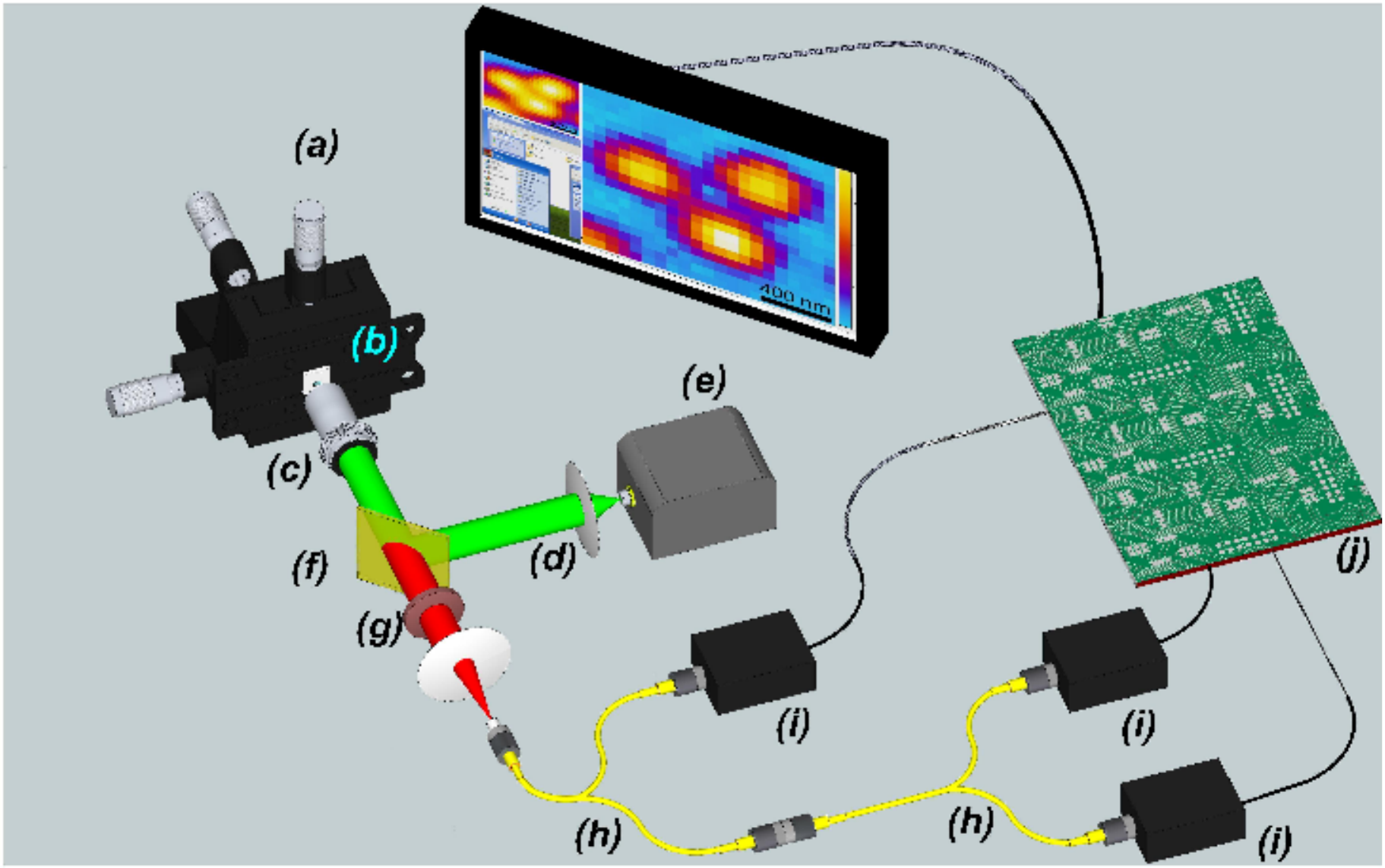}
   \end{tabular}
   \end{center}
   \caption[setup]
   {Setup of the experiment: (a) XYZ closed-loop piezo-electric stage; (b) Sample; (c) 100$\times$ oil objective; (d) excitation light (532 nm); (e) laser source (f) dichroic filter;(g) long-pass filters; (h) 50:50 fiber beam-splitter; (i) Single-photon detectors; (j) coincidence electronics. \label{fig:setup}}
   \end{figure}

Fig. \ref{fig:setup} shows the setup used for our experiment, i.e. a laser scanning single-photon sensitive confocal microscope.
The excitation light emitted by a solid state laser at 532 $nm$, coupled into a single mode fibre is collimated by a $4\times$ objective.
A dichroic mirror (Long-pass at 570 nm)  reflects the excitation light (3 mW maximum) inside the oil immersion objective (Olympus, $100\times$, NA = 1.3
) focusing inside the sample and transmits the fluorescence light towards the detecting apparatus. The sample (Element Six$^{TM}$ electronic-grade Polycrystalline CVD diamond) is mounted on a closed-loop XYZ piezo-electric stage, remotely controlled via PC, allowing nanometric-resolution positioning in a $80\mu m \times 80 \mu m$ area range. The fluorescence light (occurring within a  $640 - 800$ $nm$ spectral window) is collected by the same objective used for excitation and then passes through the dichroic mirror and a long-pass filter in order to obtain a suitable attenuation ($>10^{12}$) of the pump component. The signal is then focussed by a $f = 100$ $mm$ achromatic doublet and coupled to a $50$ $\mu m$ multimode optical fibre that not only delivers the signal to the detectors, but also acts as a pinhole for the confocal system.
The fiber leads to a detector-tree configuration \cite{tree,alan} realized by means of two integrated 50:50 beam-splitters in cascade connecting to three Single Photon Avalanche Photo-diodes (Perkin-Elmer SPCM-AQR), operating in Geiger mode. This configuration, reproducing a generalized version of the  "Hanbury-Brown and Twiss" interferometer (HBT) \cite{hbt}, allows the detection of two-/three-fold coincidences and the direct sampling of the values of the second order ($g^{(2)}$) and third order ($g^{(3)}$) \cite{elizabeth,highorderg,highorderg2} autocorrelation  functions \cite{footnote2}. The signal counts and coincidences are measured via a picosecond time-tagging module (PicoQuant Hydra-Harp).





\begin{figure}
   \begin{center}
   \begin{tabular}{c}
   \includegraphics[scale=0.4]{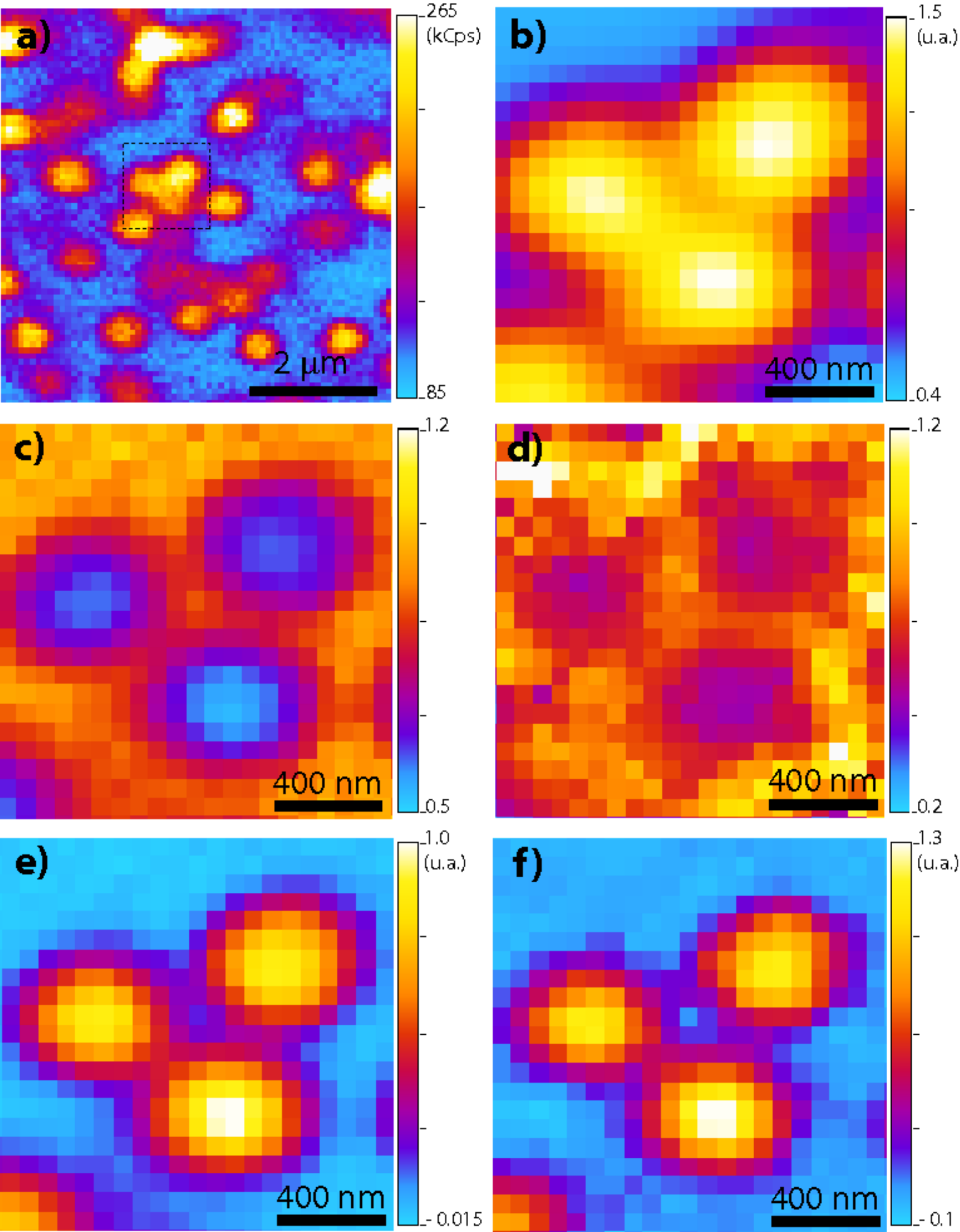}
   \end{tabular}
   \end{center}
   \caption[Tre_centri]
   {Example of the super-resolution technique applied to a cluster of 3 NV centres. a) Typical scan on a region of the sample obtained collecting the signals emitted by each centre on a pixel-by-pixel basis via single-photon sensitive confocal microscope. b) Magnification of the area of interest. c) Map of $g^{(2)}$ function. d) Map of $g^{(3)}$ function. e) Super-resolved map for {$k=2$}. f) Super-resolved map for {$k=3$}. \label{fig:Tre_centri}}
   \end{figure}

The most significant experimental results, demonstrating this method, are shown in Fig. \ref{fig:Tre_centri}. In the first inset ($a$), a typical photoluminescence map of an area of the sample obtained via our confocal microscope is shown. One can observe that in some cases the centres are well separated, while in other cases they are too close to each other to be resolved by acquiring only the fluorescence signals. For instance,  in the enlarged picture ($b$) a cluster of centres is shown that can be barely recognized as a unresolved group of three emitters. In the subsequent two pictures ($c$, $d$) the 
 maps  of (respectively) $g^{(2)}$ and $g^{(3)}$ functions are shown. Here the presence of three resolved NV centres is evident from the low value of $g^{(2)}$ and $g^{(3)}$ in correspondence of the centre's positions, while in the surrounding region the values reach $1$ because of background fluorescence light.
Finally the super-resolved maps for $k=2$  as in Eq. (\ref{eq:k2}) and $k=3$  as in Eq. (\ref{eq:k3}) are  reported in the insets ($e$) and ($f$). The progressive increase of the resolution in the above-mentioned maps  for increasing values of $k$ can be evaluated by comparing the PSF values  (360 \textpm 30 nm and 290 \textpm 30 nm) with the original resolution of the microscope (500 \textpm 16 nm), thus confirming that the resolution scales with $\sqrt{k}$.

\begin{figure}
   \begin{center}
   \begin{tabular}{c}
   \includegraphics[scale=0.45]{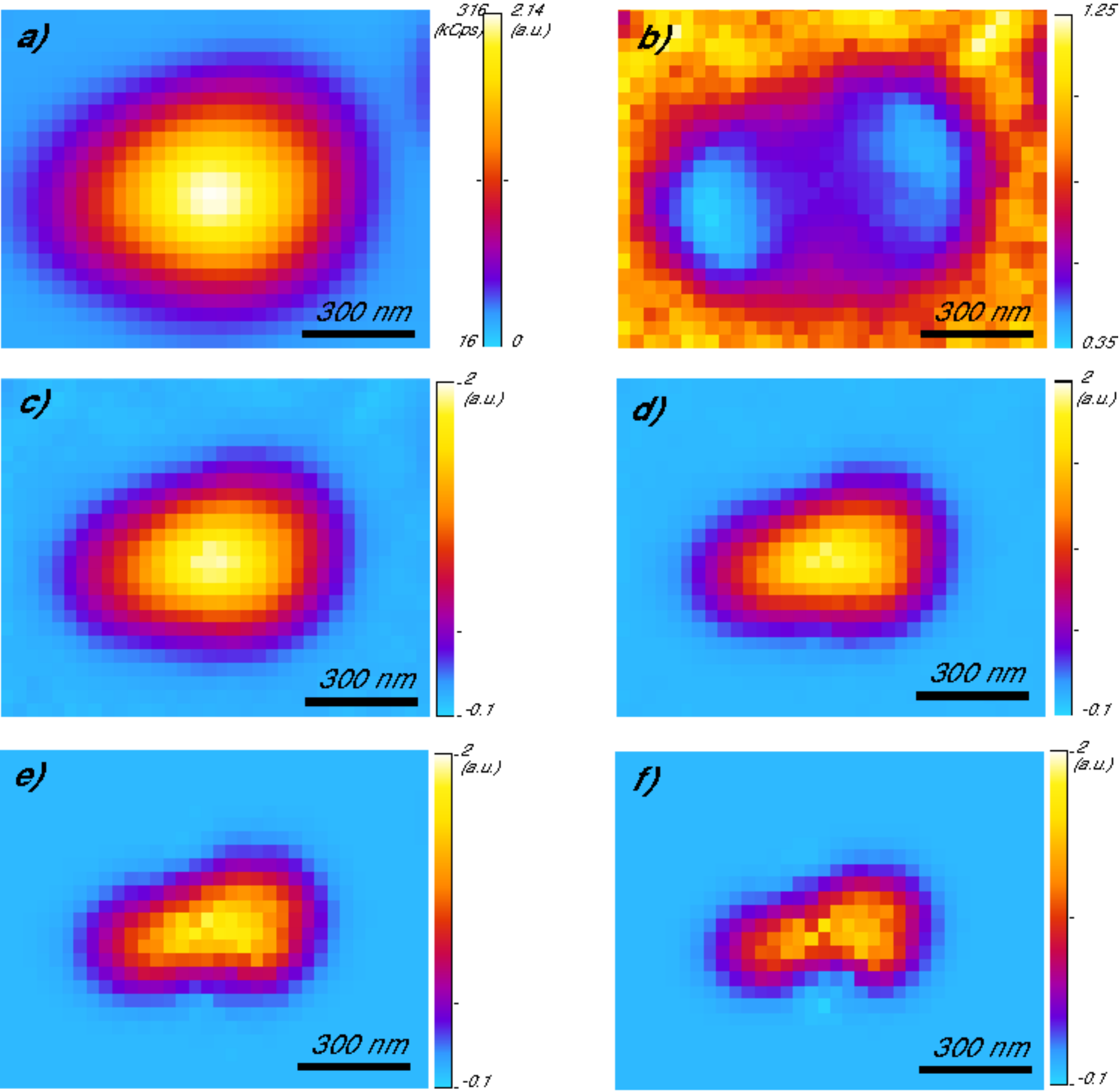}
   \end{tabular}
   \end{center}
   \caption[2_centri]
   {a) Direct mapping of the signal emitted by two NV centres, whose distance is below the FWHM of each peak, by the confocal microscope. b) Map of the $g^{(2)}$ function. c) Map obtained via the super-resolution function for $k=2$. d) Map obtained via the super-resolution function for $k=3$. e,f) Maps derived, respectively, from Eq. \ref{eq:k4} and Eq. \ref{eq:k5} for $g^{(k)} = 0$ for $k \geq 3$.\label{fig:2_centri}}
   \end{figure}

In Fig. \ref{fig:2_centri}, another example is considered:  the photoluminescence signal from the observed region of the sample is mapped on a single-peaked spot (inset $a$) whose oval shape (with a major axis larger than 500 $nm$) hints at the presence of more than one centre, although no information on the quantity and relative position of these emitters can be extracted. Since the centres are very close to each other, even if the direct scanning of $g^{(2)}$ (inset $b$) reveals the presence of two dips (i. e. two emitters), the super-resolved map obtained for $k=2$ (inset $c$) is not able to separate them. No decisive improvement of the resolution of the image can be obtained by applying the third order formula (see Eq. \ref{eq:k3}), since only two emitters are present and the  $g^{(3)}$ contribution is null everywhere, excluding background contribution that can be removed in the $g^{(2)}$ map \cite{Brouri}
. Nonetheless, in this kind of scenario, a further improvement on the resolution can be achieved by applying the series of Eq. (\ref{eq:k8}) at higher orders of $k$ with $g^{(k)} = 0$ for $k \geq 3$.



As shown in the progression of the insets $d$, $e$ and $f$ of Fig. \ref{fig:2_centri}  the resolution increases at increasing $k$ orders (respectively the third, fourth and fifth) and eventually the positions of the two centres (or their distance) can be inferred with higher precision ( 270 \textpm 70 $nm$).

In conclusion, we have experimentally demonstrated super-resolved optical imaging of NV centres in bulk single-crystal artificial diamond obtained by exploiting a technique based on the direct sampling of $k^{th}$-order autocorrelation function $g^{(k)}$ on a pixel-by-pixel basis (in particular for $k=2,3$). The analysis has been performed by feeding the signals acquired by our confocal microscope to single-photon detectors in a tree configuration. Our results show good agreement with the theoretical expectations of a narrowing of the PSF proportional to the square-root of the highest order of the  measured autocorrelation function, demonstrating the advantage of this technique and paving the way for its use in several different experimental configurations, particularly in fields where the application of STED/GSD techniques are limited. 
Furthermore, we note that the exploitation of the proposed technique is extremely straightforward to be implemented in microscopy systems investigating single-photon emitters, since they typically already make use of HBT interferometers for the $g^{(2)}$ measurements. To further enhance the resolution by measuring higher-order $g$-function it is just necessary to increase the number of ports of the HBT interferometer, namely the detector-tree, a rather simple task due to the easy scalability of these detection systems \cite{alan}. Implementing a multi-port HBT interferometer is a useful not only for increasing the image resolution, but also because it has been proven to be a powerful diagnostic tool for quantum sources, not only single-photon sources. Indeed it was proven that, by measuring $g^{(k)}$ with $k\geq2$ one can reconstruct the mean number of photons of the different quantum optical modes of a quantum field \cite{elizabeth}.
For example, the optical modes of unwanted sources of background light super-imposed to the emission of the single-photon source of interest can be easily identified \cite{elizabeth}. This can be of great interest for understanding the origin of this background and thus to find a way to eliminate it.


This research activity was supported by the following projects: FIRB project D11J11000450001 funded by MIUR; EMRP project "EXL02-SIQUTE"; "A.Di.NTech." project D15E13000130003 funded by University of Torino and Compagnia di San Paolo; NATO SPS Project 984397; "Compagnia di San Paolo" project "Beyond classical limits in measurements by means of quantum correlations".


\section*{Appendix}
In the following we present the examples of the derivation of super-resolved imaging function for $k = 2,3$ (in this section the $x$ dependance of 
\begin{itemize}
\item case $k = 2$:
The square of the signal is:
\begin{equation}
\langle\hat{N}\rangle^2=\sum_{\alpha=1}^n (\mathcal{P_\alpha})^2+\sum_{\alpha\neq\alpha '}\mathcal{P_\alpha}\mathcal{P_{\alpha '}}
\end{equation}
We would like to remove the contribution of the term  $\sum_{\alpha\neq\alpha '}\mathcal{P_\alpha}\mathcal{P_{\alpha '}}$using the expression of 
\begin{eqnarray*}
\langle\hat{N}^2\rangle=&\langle\sum_{\alpha=1}^n \hat{a}_\alpha^\dagger \hat{a}_\alpha\cdot\sum_{\alpha '=1}^n \hat{a}_{\alpha '}^\dagger\hat{a}_{\alpha '} \rangle=&\\
=&\sum_{\alpha \neq \alpha '}\langle \hat{a}_\alpha^\dagger \hat{a}_\alpha\cdot \hat{a}_{\alpha '}^\dagger\hat{a}_{\alpha '}\rangle+\sum_\alpha \langle\hat{a}_\alpha^\dagger \hat{a}_\alpha\rangle=&\\
=&\sum_{\alpha \neq \alpha '}\mathcal{P_\alpha}\mathcal{P_{\alpha '}}+\sum_\alpha \mathcal{P_\alpha}
\end{eqnarray*}
to write the second-order auto-correlation function in terms of $\mathcal{P}(x)$:
\begin{equation}
g^{(2)}=\frac{\langle N(N-1) \rangle}{\langle N \rangle^2}=\frac{\sum_{\alpha \neq \alpha '}\mathcal{P_\alpha}\mathcal{P_{\alpha '}}}{\langle N \rangle^2}
\end{equation}
This allows to obtain a map of the quantity:
\begin{equation}
\sum_{\alpha=1}^n (\mathcal{P_\alpha})^2=\langle N \rangle^2(1-g^{(2)})
\end{equation}
that is the super-resolved image.
\item case $k = 3$:
The third power of the signal can be written as:
\begin{eqnarray*}
\langle\hat{N}\rangle^3=&\sum_{\alpha=1}^n (\mathcal{P_\alpha})^3+3\sum_{\alpha\neq\alpha '}(\mathcal{P_\alpha})^2\mathcal{P_{\alpha '}}+\\+&\sum_{\alpha\neq\alpha '\neq\alpha ''}\mathcal{P_\alpha}\mathcal{P_{\alpha '}}\mathcal{P_{\alpha ''}}
\end{eqnarray*}
Since:
\begin{equation}
g^{(3)}=\frac{\langle N(N-1)(N-2) \rangle}{\langle N \rangle^3}=\frac{\sum_{\alpha\neq\alpha '\neq\alpha ''}\mathcal{P_\alpha}\mathcal{P_{\alpha '}}\mathcal{P_{\alpha ''}}}{\langle N \rangle^3}
\end{equation}
we observe that:
\begin{eqnarray*}
\sum_{\alpha=1}^n \mathcal{P_\alpha}[\sum_{\alpha\neq\alpha '}\mathcal{P_\alpha}\mathcal{P_{\alpha '}}]=&2\sum_{\alpha\neq\alpha '}(\mathcal{P_\alpha})^2\mathcal{P_{\alpha '}}+\\+&\sum_{\alpha\neq\alpha '\neq\alpha ''}\mathcal{P_\alpha}\mathcal{P_{\alpha '}}\mathcal{P_{\alpha ''}}
\end{eqnarray*}
So the super-resolved image at third-order is:
\begin{equation}
\sum_{\alpha=1}^n (\mathcal{P_\alpha})^3=\langle N \rangle^3(1-\frac{3}{2}g^{(2)}+\frac{1}{2}g^{(3)})
\end{equation}
\end{itemize}



\begin{thebibliography}{99}
\bibitem{footnote1} The whole treatment holds also in Continuous Wave (CW) excitation, considering  $d\mathcal{P}(x)$ as the infinitesimal probability of detecting the photon within an infinitesimal  time interval dt (In this case the relation will be: $d\mathcal{P} = I dt$  where $I$ is the photon emission intensity , measured as detected photons/s).
\bibitem{footnote2} Note that the lack of balance of the splitting ratio among the different ports of the HBT is not a problem in this case since the (higher order) correlation functions $g^{(k)}$ are independent on the "effective" quantum efficiency \cite{alan}.
\end{thebibliography}
\end{document}